\documentclass{ws-procs9x6}

\begin{document}

\title{PRODUCTION OF NEUTRON-RICH LAMBDA HYPERNUCLEI AT J-PARC}

\author{A. SAKAGUCHI$^a$, S. AJIMURA$^b$, H. BHANG$^c$, L. BUSSO$^{d,e}$, 
M. ENDO$^a$, D.~FASO$^{d,e}$, T. FUKUDA$^f$, T. KISHIMOTO$^a$, K. MATSUDA$^a$, 
K. MATSUOKA$^a$, Y.~MIZOI$^f$, O.~MORRA$^{e,g}$, H. NOUMI$^b$, P. K. SAHA$^h$, 
C. SAMANTA$^{i,j}$, Y.~SHIMIZU$^a$, T.~TAKAHASHI$^k$, T. N. TAKAHASHI$^\ell$ 
and K. YOSHIDA$^a$}

\vspace{2mm}

\address{
$^a$Department of Physics, Osaka University, 
Toyonaka, Osaka 560-0043, Japan \\
$^b$Research Center for Nuclear Physics, Osaka University, \\
Ibaraki, Osaka 567-0047, Japan \\
$^c$Department of Physics, Seoul National University, 
Kwanak-gu, Seoul 151-747, Korea \\
$^d$Dipartimento di Fisica Generale, Universit\`{a} di Torino, \\
via Pietro Giuria, I-10125 Torino, Italy \\
$^e$Istituto Nazionale Fisica Nucleare, Sezione di Torino, \\
via Pietro Giuria, I-10125 Torino, Italy \\
$^f$Department of Engineering, Osaka Electro-Communication University, \\
Neyagawa, Osaka 572-8530, Japan \\
$^g$Istituto di Fisica dello spazio Interplanetario, \\
Corso Fiume 4, 10133 Torino, Itary \\
$^h$Japan Atomic Energy Agency, 
Tokai, Ibaraki 319-1195, Japan \\
$^i$Physics Department, University of Richmond, 
Richmond, VA 23173, USA \\
$^j$Saha Institute of Nuclear Physics, 
1/AF Bidhannagar, Kolkata 700064, India \\
$^k$High Energy Accelerator Research Organization, 
Tsukuba, Ibaraki 305-0801, Japan \\
$^\ell$Department of Physics, University of Tokyo, 
Bunkyo-ku, Tokyo 113-8654, Japan}

\begin{abstract}
We discuss the usefulness of the double charge-exchange reactions 
(DCX) for the production of the neutron-rich $\Lambda$-hypernuclei.
We believe the $(\pi^-,K^+)$ reaction is one of the most promising 
DCX reactions, and propose to produce the neutron-rich 
$\Lambda$-hypernuclei, $^6_\Lambda$H and $^9_\Lambda$He, at the 
J-PARC 50 GeV PS by the reaction (J-PARC E10 experiment).
The design of the experiment is presented.  
\end{abstract}


\bodymatter


\section{Physics Motivation}

We are preparing the J-PARC E10 experiment for the study of the 
neutron-rich $\Lambda$-hypernuclei at the J-PARC 50~GeV PS facility.
In this paper, we discuss the issues and the design of the experiment.


\subsection{Spectroscopy of $\Lambda$-hypernuclei and new tools}

The $\Lambda$-hypernucleus was identified experimentally for the first 
time in 1953 in a nuclear emulsion exposed to cosmic 
rays\cite{danysz53}.  
Since then, number of experiments have been carried out, innovative 
methods/techniques have been developed and many aspects of the 
$\Lambda$-hypernuclei have become clear.  
One of the important subjects of the studies of the $\Lambda$-hypernuclei 
in the past was the precise measurements of the level structures of the 
$\Lambda$-hypernuclei, that made it possible to discuss the underlying 
hyperon-nucleon strong interaction.  
The similarity of the $\Lambda$~hyperon with nucleon is one of the key 
properties which brings the rich spectra of the $\Lambda$-hypernuclei.  
Another important property is the additional binding due to the 
$\Lambda$ hyperon, so we expect the {\em hypernuclear chart} is even 
richer than the ordinary {\em nuclear chart}\cite{samanta08}.  

\begin{figure}
\begin{center}
\psfig{file=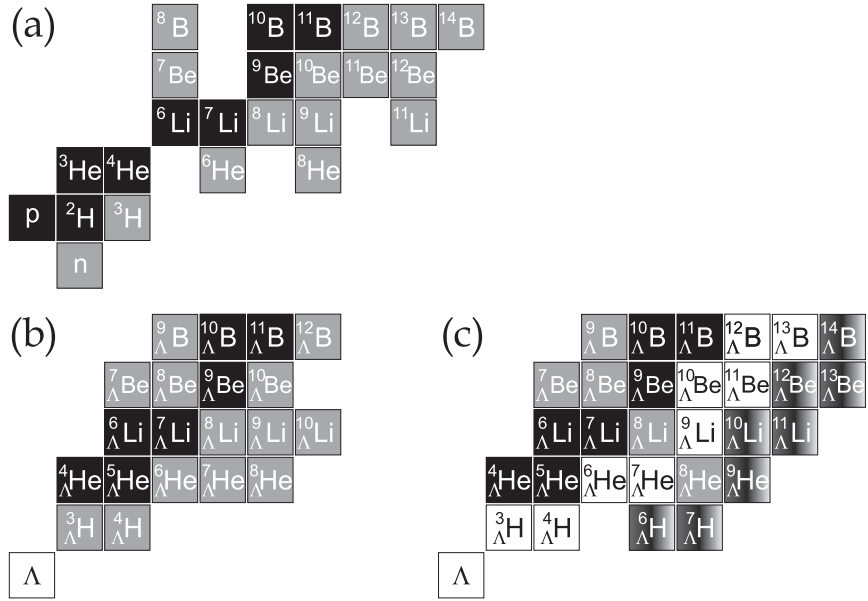,width=4.5in}
\end{center}
\caption{(a) chart of light nuclei. 
(b) chart of $\Lambda$-hypernuclei ever produced. 
(c) boxes with the graduation correspond to hypernuclei to be 
produce via the DCX reactions.
See text for more details.}
\label{fig:chart}
\end{figure}

On the other hand, we have surveyed only a small fraction of 
hypernuclei in the {\em hypernuclear chart}.  
One of reasons of the limited survey was that we mainly used 
the $(K^-,\pi^-)$ and the $(\pi^+,K^+)$ reactions to produce 
the $\Lambda$-hypernuclei.  
\Fref{fig:chart}(b) shows hypernuclei so far identified; 
black colored ones were directly produced via the $(K^-,\pi^-)$ 
and the $(\pi^+,K^+)$ reactions on stable nuclear targets 
and gray colored 
ones were observed as hyperfragments in the nuclear emulsion 
experiments.
The chart looks already compatible with that of the 
ordinary nuclei (\fref{fig:chart}(a)), but the information on the 
hyperfragments from the nuclear emulsion experiments were quite 
restricted. 

To survey wider area of the hypernuclear chart in further detail, 
we need new spectroscopic tools. 
If we employ the charge exchange reactions, we can directly produce 
many neutron-rich $\Lambda$-hypernuclei as shown in 
\fref{fig:chart}(c); hypernuclei in the white boxes are produced 
by the single charge-exchange reactions and those in the boxes 
with the graduation are produced by the double charge-exchange 
reactions (DCX), such as the $(K^-,\pi^+)$ and $(\pi^-,K^+)$ 
reactions.


\subsection{$\Lambda$N interaction and neutron-rich $\Lambda$-hypernuclei}

One of interesting aspects of the $\Lambda$N interaction in the 
hypernuclei is the phenomenon so-called $\Lambda$N-$\Sigma$N mixing.
The mass difference between the $\Lambda$ and $\Sigma$ hyperons 
is small, $m_\Sigma-m_\Lambda\sim$77MeV/c$^2$, compared with that 
of the nucleon and $\Delta$ isobar, $m_\Delta-m_N\sim$290MeV/c$^2$.
This situation makes the effect of the $\Lambda$N-$\Sigma$N mixing 
quite important in the hypernuclear level structure\cite{gibson72}.  
The strong $\Lambda$N-$\Sigma$N mixing introduces an additional 
effective two-body $\Lambda$N interaction and also the three-body 
interaction among the $\Lambda$NN subsystem in a 
$\Lambda$-hypernucleus as recently discussed by Akaishi 
{\it et~al.}\cite{akaishi00}. 
The additional interaction due to the $\Lambda$N-$\Sigma$N mixing 
is believed to increase the $\Lambda$N attraction, and the attractive 
interaction may affect to the fraction of hyperons and EOS of the 
matter in the core of the neutron stars.  

Since the $\Lambda$ and $\Sigma$ hyperons have different isospins, 
I=0 and 1 respectively, the $\Lambda$N-$\Sigma$N coupling may be 
large only for hypernuclei with non-zero isospin due to the isospin 
conservation.  
We also expect the mixing effect is significant in the neutron-rich 
$\Lambda$-hypernuclei which have large values of the isospin.   
This mixing effect manifests itself in a $\Sigma$ component 
of the $\Lambda$-hypernuclear states, which can be useful to produce 
the neutron-rich $\Lambda$-hypernuclei by the $(\pi^-,K^+)$ reaction 
(see \sref{sec:reaction} for more details).  


\subsection{Production of neutron-rich $\Lambda$-hypernuclei}

A pilot experiment attempted to produce $\Lambda$-hypernuclei 
away from the stability-line was performed at KEK-PS by using the 
$(K^-_{Stopped},\pi^+)$ reaction\cite{kubota96}.  
In the experiment, only upper limits were obtained for the 
production rates of the neutron-rich $\Lambda$-hypernuclei 
($^6_\Lambda$He, $^{12}_{~\Lambda}$Be and $^{16}_{~\Lambda}$C) 
due to tiny 
branching ratios to the DCX channel and a huge background from the 
in-flight hyperon decays, $\Sigma^+$$\rightarrow$$n\pi^+$.  
An improved study with the $(K^-_{Stopped},\pi^+)$ reaction has 
been carried out for the $^{6}_\Lambda$H, $^{7}_\Lambda$H and 
$^{12}_\Lambda$Be hypernuclei by the FINUDA collaboration at 
Frascati-DA$\Phi$NE\cite{agnello05}, but the clear identification 
of the production of the neutron-rich $\Lambda$-hypernuclei was 
not accomplished. 
Another experiment to produce not only the neutron-rich but also 
the proton-rich $\Lambda$-hypernuclei is in preparation by using 
the relativistic heavy ion beams at GSI\cite{saito08}.

The other promising DCX reaction to produce the neutron-rich 
$\Lambda$-hypernuclei is the $(\pi^-,K^+)$ reaction.
A neutron-rich $\Lambda$-hypernucleus, $^{10}_{~\Lambda}$Li, was 
attempted to produce at KEK-PS by the $(\pi^-,K^+)$ reaction 
with the 1.05 and 1.2 GeV/c pion beams (KEK-E521 experiment)\cite{saha05}.  
In the experiment, clear signal events were observed in the $\Lambda$ 
bound region in the missing mass spectrum of the 
$^{10}{\rm B}(\pi^-,K^+)$ reaction.  
Although the production cross section of the $^{10}_{~\Lambda}$Li 
hypernucleus was estimated to be very small ($\sim$10nb/sr), roughly 
$10^{-3}$ of that of the $(\pi^-,K^+)$ reaction (typically 
10$\mu$b/sr), the experimental data may provide new 
information on the structure of the $\Lambda$-hypernuclei with a 
large number of excess neutrons.  
Compared with the $(K^-_{Stopped},\pi^+)$ reaction, the $(\pi^-,K^+)$ 
reaction is almost background free at the $\Lambda$ bound region.  


\subsection{Reaction mechanism and $\Lambda$N-$\Sigma$N mixing}
\label{sec:reaction}

The KEK-E521 experiment reported that the production cross 
sections of the $^{10}_{~\Lambda}$Li hypernucleus by the 
$(\pi^-,K^+)$ reaction were 5.8~nb/sr 
and 11.3~nb/sr at the pion beam energies of 1.05~GeV/c and 
1.20~GeV/c, respectively.  
Theoretical calculations based on the two-step reaction mechanism, 
$\pi^-pp \to K^0\Lambda p \to K^+\Lambda n$ or 
$\pi^-pp \to \pi^0np \to K^+n\Lambda$, 
failed to reproduce the beam momentum dependence of the production 
cross sections.  

Another theoretical treatment of the DCX reaction is the single-step 
reaction mechanism.  
In the reaction mechanism, the single-step $\pi^-p \to K^+\Sigma^-$ 
reaction occurs, and the produced $\Sigma^-$ is converted to 
$\Lambda$ through the $\Lambda$N-$\Sigma$N mixing in the 
hypernucleus.  
Since the threshold of the $\pi^-p \to K^+\Sigma^-$ reaction is 
around 1.045 GeV/c, the single-step calculations predict 
the production cross sections consistent with the experiment. 
Recent results of the single-step calculations indicated the 
importance of the mixing effect and also the effect of the 
intermediate $\Sigma$-nucleus interaction to the $(\pi^-,K^+)$ 
reaction cross sections\cite{tretyakova01,harada09}.


\section{Design of J-PARC E10 Experiment}

The J-PARC E10 experiment was proposed to utilize the $(\pi^-,K^+)$ 
reaction to produce the new neutron-rich $\Lambda$-hypernuclei, 
$^{6}_\Lambda$H and $^{9}_\Lambda$He.  
We are proposing to use the K1.8 secondary beam line in the hadron 
experimental hall of the J-PARC 50 GeV proton synchrotron (50 GeV PS) 
facility.  
The 50 GeV PS has already started to accelerate proton beams and the 
K1.8 beam line is to be constructed in 2009.  

The most important issues of the E10 experiment are the handling of 
the high-intensity pion beams and the efficient measurement of the 
produced kaons to override the tiny production cross section 
of the DCX reaction.  
We are preparing new tracking detectors in the beam line spectrometer 
of the K1.8 beam line to accept the high intensity pion beams, about 
1.5$\times$$10^7~\pi^-/spill$ with 3~s beam spill.  
The produced positive kaons are detected by the Superconducting Kaon 
Spectrometer (SKS)\cite{sks95} to be moved from KEK to the J-PARC site.  
SKS has a large geometrical acceptance, about 100 msr, together with 
the good momentum resolution, $\Delta p/p \sim 10^{-3}$(FWHM).  
The practical overall mass resolution, roughly 2.5 MeV/c$^2$(FWHM),  
of the hypernuclei calculated by the missing mass method mainly 
comes from the contribution of 
the energy loss straggling of $\pi^-$ and $K^+$ in the relatively 
thick nuclear target (about $3.5~g/cm^2$).  
We estimated the yield of the $^{9}_\Lambda$He hypernucleus by 
employing the additional parameters listed in \tref{tbl:parameter}, 
and the estimated yield were about 300 events during 3 weeks of 
beamtime.  
\begin{table}
\tbl{Additional parameters used for the $^9_\Lambda$He yield 
estimation.}
{\begin{tabular}{cccc}\toprule
$\pi^-$ momentum & PS acceleration cycle & $K^+$ decay factor & Analysis efficiency \\ \colrule
1.2 GeV/c & 5.7 s & 0.5 & 0.5 \\ \botrule
\end{tabular}}
\label{tbl:parameter}
\end{table}
The number of events is not so many, but is about 7 times larger 
than that of the KEK-E521 experiment (yield was about 47 events).  

\begin{figure}
\begin{center}
\psfig{file=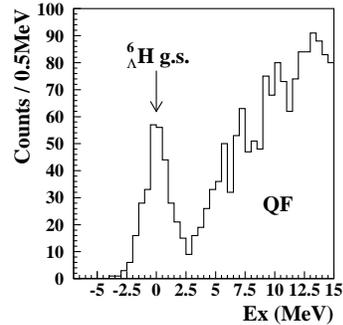,width=1.8in}
\end{center}
\caption{The excitation energy spectrum of the $^6_\Lambda$H 
hypernucleus estimated by a simulation calculation.
See text for more details.}
\label{fig:spectrum}
\end{figure}

\Fref{fig:spectrum} shows the excitation energy spectrum of 
the $^6_\Lambda$H hypernucleus estimated by a simulation calculation.  
The overall excitation energy resolution of 2.5 MeV(FWHM) and 
the $^6_\Lambda$H yield of 300 events were assumed.
The figure tells the peak which corresponds to the ground state 
of the $^6_\Lambda$H hypernucleus is clearly separated from the 
continuum of the quasi-free $\Lambda$ production process thanks 
to the good energy resolutions of the beam line spectrometer 
and SKS.  
The statistical error is small enough to determine the binding 
energy of the neutron-rich $\Lambda$-hypernucleus with a 
resolution down to 0.1~MeV (rms).  

We are also preparing the upgrade of the detectors in SKS to make 
the momentum acceptance wider.  
The upgrade may enable us to measure the $(\pi^-,K^+)$ reaction in 
the $\Lambda$ and the $\Sigma^-$ production regions at the same time, 
and the wide momentum acceptance is quite useful to make precise 
calibrations of the absolute scale of the binding energy of the 
hypernuclei and to monitor the stability of the spectrometer systems.  

We hope we will be able to start the experiment in the fiscal year 
2009.  


\end{document}